\documentclass[final,5p,times,twocolumn]{elsarticle}

\usepackage{amssymb}

\journal{Physica C}

\begin{document}

\begin{frontmatter}



\title{Shift of the maxima of the critical currents of different
 polarity relative to the zero magnetic flux along the flux axis in a superconducting asymmetric aluminum
 ring}


\author{V.~I.~Kuznetsov \corref{cor1}}
\ead{kuznetcvova@mail.ru}
\cortext[cor1]{corresponding author}
\author{O.~V.~Trofimov}
\address{Institute of Microelectronics Technology and High
Purity Materials, Russian Academy of Sciences, Chernogolovka,
Moscow Region 142432, Russia}

\begin{abstract}
We measured the rectification of an ac voltage in a structure of
superconducting circularly-asymmetric aluminum rings in series,
permeated with a magnetic flux and biased with a low-frequency
alternating current (without a dc component). This rectification
is due to the shift of the maxima of the critical currents of
different polarity relative to the zero flux in opposite
directions along the flux axis in the asymmetric ring. For the
first time, we propose a model for a temperature-dependent phase
shift equal to difference between dimensionless kinetic
inductances of wide and narrow semirings having the same length
and thickness. The shift is not zero in the case of different
critical currents densities in both semirings. This is possible
only in a situation of different critical temperatures of both
semirings. The model describes well the temperature-dependent
shift of the maxima of the critical currents, answers the
long-standing mysterious challenge of the shift and removes
extremely strange contradiction between the results of different
measurements, previously found in circularly-asymmetric aluminum
structures.
\end{abstract}



\begin{keyword}
circularly-asymmetric aluminum ring \sep superconducting critical
temperature \sep rectified magnetic-field-dependent voltage \sep
magnetic-field-dependent critical current \sep Josephson critical
current \sep kinetic inductance

\end{keyword}

\end{frontmatter}



\section{Introduction}

It is well known that due to the requirement to quantize the
superconducting fluxoid, a circulating magnetic-field-dependent
oscillating current $I_{r}(\Phi/\Phi_{0})$ arises in a
superconducting thin-walled hollow cylinder (ring) pierced with a
magnetic flux $\Phi$ near the critical temperature $T_{c}$
\cite{tinkham}. This circulating current changes with a period in
the magnetic field $dB$ corresponding to the superconducting
quantum of the magnetic flux $\Phi_{0}=dBS_{eff}$ through the
effective ring area $S_{eff}$. This current $I_{r}(\Phi/\Phi_{0})$
as a function of the normalized flux $\Phi/\Phi_{0}$ is directly
proportional to the superconducting velocity
$v_{s}(\Phi/\Phi_{0})$. The circulating current is given by
$I_{r}(\Phi/\Phi_{0})=-2I_{rm}(n-\Phi/\Phi_{0})$ when
$\Phi/\Phi_{0}$ varies from $n-0.5$ to $n+0.5$, where $n$ is an
integer. This current has breaks (jumps), reaching its extrema
$+I_{rm}$ and $-I_{rm}$ simultaneously at $\Phi/\Phi_{0}=n+0.5$,
where two oppositely directed circulating currents of the same
magnitude have the same probability (Fig. \ref{f1}, upper inset).
The current is alternating-sign, changing its sign at
$\Phi/\Phi_{0}=n$ and $\Phi/\Phi_{0}=n+0.5$.

Oscillations of dc voltage $V_{dc}(\Phi/\Phi_{0})$ (oscillations
of resistance $dR(\Phi/\Phi_{0})$ caused by oscillations of the
critical temperature $dT_{c}(\Phi/\Phi_{0})$ with the period
$\Phi_{0}$) arise in a superconducting hollow cylinder (ring) of
small diameter permeated with a magnetic flux, when through it a
small direct current $I_{dc}$, close to the critical current
$I_{c}$, is passed at temperatures $T$ close to $T_{c}$ (the
Little-Parks effect \cite{tinkham, little}). The change in
resistance $dR(\Phi/\Phi_{0})$ is directly proportional to
$v_{s}^{2}(\Phi/\Phi_{0})$ and has minima at $\Phi/\Phi_{0}=n$ and
maxima at $\Phi/\Phi_{0}=n+0.5$.

An oscillating rectified time-averaged direct voltage
$V_{rec}(\Phi/\Phi_{0})$ with the period $\Phi_{0}$ appears in a
superconducting circularly-asymmetric aluminum ring and such rings
in series, placed in a magnetic field and biased with a sinusoidal
alternating current (without a dc component)
$I_{ac}(t)=I_{acm}sin(2\pi \nu t)$ (where $\nu$ is the ac
frequency, $I_{acm}$ is the ac amplitude close to $I_{c}$), at $T$
close to $T_{c}$ \cite{dubjetplet03, kuznprb08, kuznphysica13,
karpii07, nikulov07}. An example of one such ring is shown in the
lower insert of Fig. \ref{f1}. The circular asymmetry of the ring
was usually created due to the fact that the width of the wide
semiring $w_{w}$ was twice as large as the width of the narrow
semiring $w_{n}$.

The rectified direct voltage $V_{rec}(\Phi/\Phi_{0})$ is obtained
as a result of time averaging of the ac voltage
$V_{ac}(\Phi/\Phi_{0},t)$ over the time interval $dt^{*}>20dt_{I}$
(where $dt_{I}$ is the ac period), i.e.
$V_{rec}(\Phi/\Phi_{0})=1/dt^{*}\int
^{dt^{*}}_{0}V_{ac}(\Phi/\Phi_{0},t)dt$.

The rectified voltage $V_{rec}(\Phi/\Phi_{0})$ is
alternating-sign. This voltage is zero at $\Phi/\Phi_{0}=n$ and
$n+0.5$ and has extrema lying in the interval between
$\Phi/\Phi_{0}=n$ and $n+0.5$ \cite{dubjetplet03, karpii07,
nikulov07}. The conditional positive (negative) voltage $\pm
V_{ac}(\Phi/\Phi_{0}, t)$ occurs when the combined action of the
conditionally positive (negative) half-wave of the ac $I_{ac}(t)$
and the circulating current $I_{r}(\Phi/\Phi_{0})$ leads to the
achievement of the critical current density at a certain part of
the ring.

In a circularly-symmetric ring, the condition
$I_{c+}(\Phi/\Phi_{0})=-I_{c-}(\Phi/\Phi_{0})$ is satisfied, where
both critical currents for positive and negative ac half-waves
$I_{c+}(\Phi/\Phi_{0})$ and $I_{c-}(\Phi/\Phi_{0})$ are symmetric
with respect to zero, have minima and maxima at $\Phi/\Phi_{0}=n$
and $n+0.5$, respectively. In a circularly-symmetric ring, the
voltage $V_{rec}(\Phi/\Phi_{0})$ is zero, since the conditionally
positive voltage compensates for the negative voltage.

In case of violation of circular symmetry, a nonzero voltage
$V_{rec}(\Phi/\Phi_{0})$ arises, since the critical currents for
positive and negative ac half-waves are different. Initially
\cite{dubjetplet03} it was assumed that the critical currents for
opposite directions of the applied alternating current differed in
amplitude. Further study \cite{karpii07, nikulov07} showed that
the maximum magnitudes of the critical currents for opposite
directions of the applied current were the same. In addition, the
maxima of the critical currents for opposite directions of the ac
$I_{c+}(\Phi/\Phi_{0}+\phi_{l})$ and
$I_{c-}(\Phi/\Phi_{0}-\phi_{l})$ were shifted relative to the flux
zero in different directions along the normalized flux axis by the
difference $\phi_{l}=\Delta \Phi/\Phi_{0}$, measured in fractions
of the oscillation period $\Phi_{0}$. In this case, at high
currents, the mutual shift of the maxima of the critical currents
relative to each other is equal to $2\phi_{l}$ and reaches a value
of 0.5, corresponding to half the oscillation period. In
\cite{karpii07}, the shift $2\phi_{l}$ was measured in single
aluminum rings depending on the value of the ring circular
asymmetry $w_{w}/w_{n}$.

It was found that the mutual shift $2\phi_{l}\approx 0.5$ and did
not depend on $T$ when the circular asymmetry changed in the range
$w_{w}/w_{n}=2-1.25$. Whereas the measured mutual shift
$2\phi_{l}\approx 0.36$ in a structure of eighteen
circularly-asymmetric rings in series with the ratio
$w_{w}/w_{n}=2$ \cite{nikulov07}. More accurate measurements
showed that the mutual shift $2\phi_{l}$ can be much less than 0.5
and grows with decreasing $T$ at temperatures close to $T_{c}$ (at
low critical currents) in a circularly-asymmetric single ring with
a radius of $r=2$ $\mu$m and a structure of one hundred and ten
rings in series ($r=1$ $\mu$m) \cite {gurtvoi0910}.

It turned out that the sum $I_{\Sigma
c}(\Phi/\Phi_{0})=I_{c+}(\Phi/\Phi_{0}+\phi_{l})+I_{c-}(\Phi/\Phi_{0}-\phi_{l})$
becomes zero at $\Phi/\Phi_{0}=n$ and $n+0.5$ and reaches extrema
in the interval between $\Phi/\Phi_{0}=n$ and $n.5$ (close to $n$)
\cite{karpii07, nikulov07}. The $I_{\Sigma c}(\Phi/\Phi_{0})$ sum
behaves like the rectified voltage $V_{rec}(\Phi/\Phi_{0})$. It
was found that $V_{rec}(\Phi/\Phi_{0})$ is directly proportional
to $I_{\Sigma c}(\Phi/\Phi_{0})$.

Rectification of an ac voltage $V_{ac}(\Phi/\Phi_{0},t)$ in a
superconducting circularly-asymmetric single aluminum ring or in a
structure of such rings in series, permeated with a magnetic flux
and biased with an ac $I_{ac}(t)$ (without a dc component) with an
ac amplitude $I_{acm}$ close to $I_{c}$ at temperatures close to
$T_{c}$ was measured earlier \cite{dubjetplet03, kuznprb08,
kuznphysica13, karpii07, nikulov07}.

However, to date, unexpected mutual shift $2\phi_{l}$ of the
maxima of the critical currents of opposite directions relative to
each other along the flux axis, leading to rectification, has not
found any explanation. Moreover, there are serious contradictions
between different measurements and the theoretical model
\cite{karpii07, nikulov07}. In a circularly-asymmetric ring, the
maxima and minima of the oppositely directed critical currents
$I_{c+}(\Phi/\Phi_{0}+\phi_{l})$ and
$I_{c-}(\Phi/\Phi_{0}-\phi_{l})$ are shifted by $\phi_{l}$
relative to the values of the normalized flux equal to $n$ and
$n+0.5$, and corresponding to the maxima and minima of critical
currents in a circularly-symmetric ring, respectively. At a time
when for both circularly asymmetric and symmetric rings, the
minima and maxima of the resistance oscillations
$dR(\Phi/\Phi_{0})$ at a small dc are reached at integer $n$ and
half-integer $n+0.5$.

No shift $\phi_{l}$ on the dependence $dR(\Phi/\Phi_{0})$ for a
circularly-asymmetric ring contradicts the presence of this shift
for the maxima of the critical currents
$I_{c+}(\Phi/\Phi_{0}+\phi_{l})$ and
$I_{c}(\Phi/\Phi_{0}-\phi_{l})$ in the same ring \cite{karpii07}.
In addition, the authors of  \cite{nikulov07} did not find breaks
(jumps) of the critical currents of opposite directions expected
by them, which, in their opinion, should be observed due to the
break (jump) of the circulating current $I_{r}(\Phi/\Phi_{0})$ at
values of the normalized flux equal to $n+0.5$ (Fig. \ref{f1},
upper inset).

From a practical point of view, a superconducting
circularly-asymmetric ring pierced with a magnetic flux can work
as an asymmetric quantum superconducting interference device of
micron size (micro-SQUID) \cite{barone, clarke}. In addition, a
superconducting circularly-asymmetric aluminum ring and a
structure of such rings in series is a very effective
magnetic-field-dependent ac voltage rectifier and can serve as a
highly sensitive detector of electromagnetic nonequilibrium noise
\cite{dubjetplet03, kuznprb08, kuznphysica13, karpii07,
nikulov07}.

In this work, in order to understand the shift that causes an, ac.
voltage rectification, we carried out resistive measurements of a
structure of superconducting circularly-asymmetric aluminum rings
in series. We also sketched a model to explain the shift. Using
this model, we calculated the temperature-dependent values of the
phase shift for certain superconducting circularly-asymmetric
aluminum structures measured in \cite{nikulov07, gurtvoi0910,
gurtvoi0603005, burlakov0609345}. In addition, we compared the
theoretical values of the phase shift with the experimental values
\cite{nikulov07, gurtvoi0910, gurtvoi0603005, burlakov0609345}.

\section{Results and Discussion}

\subsection{Measurement of the voltage rectification}

In this section, in order to better understand the features of an
ac voltage rectification in a superconducting
circularly-asymmetric aluminum structure, we measured the
rectification in three identical circularly-asymmetric rings in
series. Measurements were taken in a room protected from
high-frequency electromagnetic interference. To reduce the impact
of low-frequency mains interference and high-frequency noise on
structures, the use of digital devices is minimized. In addition,
some homemade analog instruments were powered by galvanic
batteries. The electrical wiring inside the cryostat was made
using twisted pairs of copper wires with a diameter of 0.15 mm.
One-kilo-ohm resistances were soldered between the end of each
wire and each current (potential) contact of the structure.

\begin{figure}
\begin{center}
\includegraphics[width=1\linewidth]{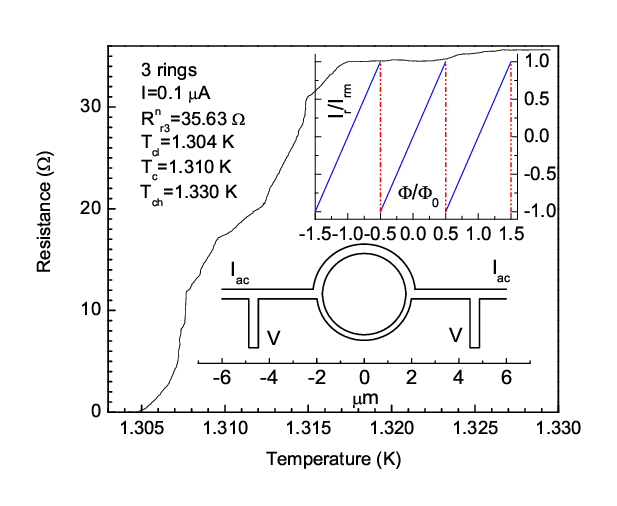}
\caption{\label{f1} (Color online) Resistive N-S transition
$R_{r3}(T)$ measured on a structure of three circularly-asymmetric
rings in series at $I_{dc}=0.1$ $\mu$A. Inset upper: expected
normalized circulating current $I_{r}/I_{rm}$ as a function of
$\Phi/\Phi_{0}$ in a superconducting circularly-symmetric ring of
small radius at $T$ close to $T_{c}$. Inset below: a sketch of one
of the rings included in the structure of three rings in series.}
\end{center}
\end{figure}

The structure of three identical circularly-asymmetric rings in
series is fabricated by thermal deposition an aluminum film with
the thickness of $d=50$ nm on a silicon substrate using the
lift-off process of electron beam lithography. One of the rings of
this structure is shown in the lower insert of Fig. \ref{f1}. The
inner radius of the ring is $r_{i}=1.76$ $\mu$m, the outer radii
of the upper and lower semirings are $r_{e1}=2.16$ $\mu$m and
$r_{e2}=2$ $\mu$m, respectively. The widths of the upper and lower
semirings are $w_{w}=0.41$ and $w_{n}=0.24$ $\mu$m, respectively.
The current and potential wires have the same width $w_{I}=0.41$
$\mu$m. The distance between potential wires of one ring is 9.34
$\mu$m. Thus, the measured voltage on one ring includes the
voltage on the ring and on the current wires.

The resistive transition $R_{r3}(T)$ from the normal (N) state to
the superconducting (S) state of a structure of three rings in
series is recorded at a dc of $\mu$A (Fig. \ref{f1}). The
resistance of the structure in the normal state is
$R^{n}_{r3}=35.63$ $\Omega$. Critical temperatures are determined
$T_{cl}=T_{cl}(0)=1.304$, $T_{c}=T_{c}(0.5)=1.310$,
$T_{ch}(0.96)=1.317$, and $T_{ch}=T_{ch}(1)=1.330$ K,
corresponding to four levels of the N-S transition of the
structure $R_{r3}(T)/R^{n}_{r3}=0$, 0.5, 0.96, and 1,
respectively. The temperature range of the N-S transition is
$T_{ch}-T_{cl}=0.0253$ K.

The resistance ratio of the structure is $R_{300}/R_{4.2} \approx
2$. The mean free path of electrons $l_{el}=9.2$ nm is calculated
using the expression \cite{ashcroft}
$l_{el}=l_{ph}(R_{300}/R_{4.2}-1)$, where $l_{ph}=9.2$ nm is the
effective electron path length for scattering by phonons for
aluminum at $T=300$ K. To calculate $l_{ph}$, we took the
experimental values of the interatomic distances and Young's
modulus, equal to $a=0.22$ nm and $Y=67$ GPa, respectively. The
structure is a dirty superconductor, since $l_{el}<<\xi_{0}=1.6$
$\mu$m, where $\xi_{0}$ is the superconducting coherence length of
pure aluminum at $T=0$. Then, near $T_{c}$, the Ginzburg-Landau
coherence length is $\xi(T)=\xi(0)(1-T/T_{c})^{-1/2}$, here
$\xi(0)=0.85(l_{el}\xi_{0})^{1/2}=0.1$ $\mu$m \cite{schmidt}.

\begin{figure}
\begin{center}
\includegraphics[width=1\linewidth]{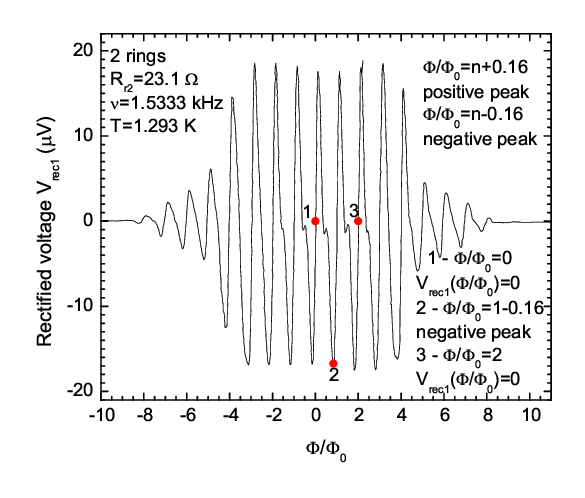}
\caption{\label{f2} (Color online) Rectified direct voltage
$V_{rec1}(\Phi/\Phi_{0})$ as a function of $\Phi/\Phi_{0}$
measured in two rings in series biased with an alternating cosine
current (without a dc component) $I_{ac1}(t)$ with the frequency
$\nu=1.5333$ kHz and the ac amplitude $I_{acm1}=3.88$ $\mu$A
greater than the critical current $I_{c}(T,B=0)$ at $T=1.293$ K.
The numbers 1, 2, and 3 are placed near the marked points of the
$V_{rec1}(\Phi/\Phi_{0})$ curve corresponding to three values
$\Phi/\Phi_{0}=0$, 1-0.16, and 2, respectively. At these points,
the oscillograms of the ac voltage $V_{ac1}(\Phi/\Phi_{0},t)$ were
taken at the same ac $I_{ac1}(t)$.}
\end{center}
\end{figure}

\begin{figure}
\begin{center}
\includegraphics[width=1\linewidth]{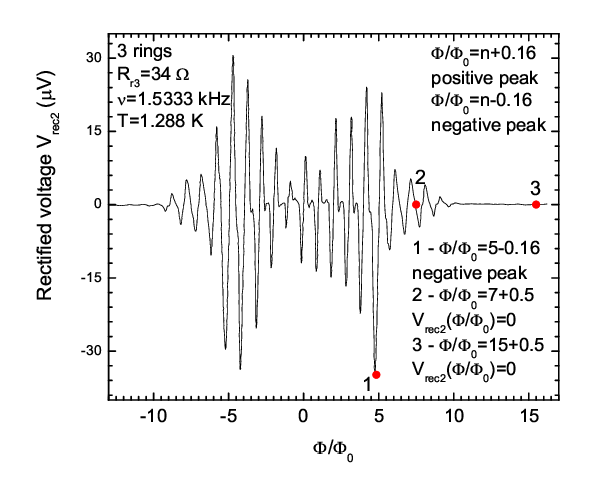}
\caption{\label{f3} (Color online) Rectified voltage
$V_{rec2}(\Phi/\Phi_{0})$ as a function of $\Phi/\Phi_{0}$,
recorded on a structure of three rings in series biased with an
alternating cosine current (without a dc component)  $I_{ac2}(t)$
with the frequency $\nu=1.5333$ kHz and the ac amplitude
$I_{acm2}=5.5$ $\mu$A less than $I_{c}(T,B=0)$ at $T=1.288$ K. The
numbers 1, 2, and 3 are placed near the marked points of the
$V_{rec2}$ curve corresponding to three values
$\Phi/\Phi_{0}=5-0.16$, 7+0.5, and 15+0.5, respectively. At these
points, the voltage oscillograms $V_{ac2}(\Phi/\Phi_{0},t)$ are
measured at the same ac $I_{ac2}(t)$.}
\end{center}
\end{figure}

Figures \ref{f2}, and \ref{f3} show two experimental curves of the
rectified time-averaged direct voltage $V_{rec1}(\Phi/\Phi_{0})$
and $V_{rec2}(\Phi/\Phi_{0})$ as functions of the normalized
magnetic flux $\Phi/\Phi_{0}$ penetrating a structure of three
circularly-asymmetric aluminum rings in series. The rectified
voltage $V_{rec1}(\Phi/\Phi_{0})$ (Fig. \ref{f2}) is taken from
two rings (with a total resistance in the normal state equal to
$R_{r2}=23.1$ $\Omega$) when passing the alternating cosine
current (without a dc component) $I_{ac1}(t)=I_{acm1}cos(2\pi \nu
t+\pi)=I_{acm1}cos(2\pi t_{norm}+\pi)$ (where the frequency
$\nu=1.5333$ kHz, the ac amplitude $I_{acm1}=3.88$ $\mu$A, here
$t_{norm}=t/dt_{I}=\nu t$ is the time normalized to the
oscillation period $dt_{I}$) at $T=1.293$ K. Voltage
$V_{rec1}(\Phi/\Phi_{0})$ is measured in the case when $I_{acm1}$
is greater than the critical current in zero magnetic field, that
is, $I_{acm1}>I_{c}(T,B=0)$. In this case, the oscillation
amplitude $V_{rec1}(\Phi/\Phi_{0})$ is maximum and almost does not
change at $-4.16<\Phi/\Phi_{0}<4.16$. These oscillations disappear
sharply at $\Phi/\Phi_{0}=-8$ and 8, despite the fact that the
wide semiring remains superconducting.

The voltage $V_{rec2}(\Phi/\Phi_{0})$ (Fig. \ref{f3}) is recorded
on three rings (with a total resistance in normal state equal to
$R_{r3}=34$ $\Omega$), biased with the alternating cosine current
(without a dc component) $I_{ac2}(t)=I_{acm2}cos(2\pi \nu
t)=I_{acm2}cos(2\pi t_{norm})$ at $T=1.288$ K, where $\nu=1.5333$
kHz, $I_{acm2}=5.5$ $\mu$A. The voltage $V_{rec2}(\Phi/\Phi_{0})$
is taken in the case when $I_{acm2}<I_{c}(T,B=0)$. In this case,
voltage oscillations are slightly weakened in low magnetic fields.
The   amplitude of $V_{rec2}(\Phi/\Phi_{0})$ oscillations
initially increases with increasing field, reaches extrema at
$\Phi/\Phi_{0} \approx -5.16$ and 5.16 (when the condition
$I_{acm2} \approx I_{c}(T,B)$ is satisfied), then decreases.
Oscillations disappear sharply at $\Phi/\Phi_{0}=-9$ and 9, while
the wide semiring remains superconducting.

The rectified voltages $V_{rec1}(\Phi/\Phi_{0})$ and
$V_{rec2}(\Phi/\Phi_{0})$ (Figs. \ref{f2}, and \ref{f3}) oscillate
with a period equal to the superconducting quantum of the magnetic
flux $\Phi_{0}=hc/2e$. The period of oscillations in the magnetic
field is $dB=1.519$ G and corresponds to the quantum of the
magnetic flux through the effective area of the ring
$S_{eff}=\Phi_{0}/dB=13.61$ $\mu$m$^{2}$. Then the effective
radius of the ring is equal to $r_{eff}=2.08$ $\mu$m and is close
to the outer radius of the lower semiring $r_{e2}=2$ $\mu$m (Fig.
\ref{f1}, lower insert).

Note that oscillations $V_{rec1}(\Phi/\Phi_{0})$ and
$V_{rec2}(\Phi/\Phi_{0})$ disappear sharply in low fields at
$\Phi/\Phi_{0}=8$ ($B=12$ G) and $\Phi/\Phi_{0}=9$ ($B=14$ G),
respectively, despite the fact that these fields do not reach
their maximum critical values for a wide semiring equal to
$B^{w}_{max}(T=1.293)=(\Phi_{0}\sqrt{3})/(\pi
\xi(T=1.293)w_{w})=31.6$ and $B^{w}_{max}(T=1.288)=36$ G for
temperatures $T = 1.293$ and 1.288 K, respectively. It is not
clear why these oscillations disappear in low fields despite the
presence of superconductivity in the ring.

It is seen that both functions $V_{rec1}(\Phi/\Phi_{0})$ and
$V_{rec2}(\Phi/\Phi_{0})$ (Figs. \ref{f2}, and \ref{f3}) have
positive and negative extrema for $\Phi/\Phi_{0}=n+\phi_{l}$ and
$n-\phi_{l}$ (where $\phi_{l}=0.16$ is a shift), respectively.
Most likely this shift $\phi_{l}=0.16$ is due to the same shift
$\phi_{l}$ of the maxima of the critical currents of different
polarity $I_{c+}(\Phi/\Phi_{0}+\phi_{l})$ and
$I_{c}(\Phi/\Phi_{0}-\phi_{l})$ relative to the zero flux in
opposite directions along the axis of the normalized magnetic
flux.

This shift corresponds to a mutual shift $2\phi_{l}=0.32$ of the
maxima of the critical currents of different polarity relative to
each other in opposite directions. Moreover, both functions become
zeros at $\Phi/\Phi_{0}=n$ and $n+1/2$. Simultaneously with the
measurements of these functions, oscillograms of the ac voltage
$V_{ac}(\Phi/\Phi_{0},t)$ are recorded at certain values of
$\Phi/\Phi_{0}$, when an alternating current (with a zero dc
component) $I_{ac}(t)$ passed through the structure.

\begin{figure}
\begin{center}
\includegraphics[width=1\linewidth]{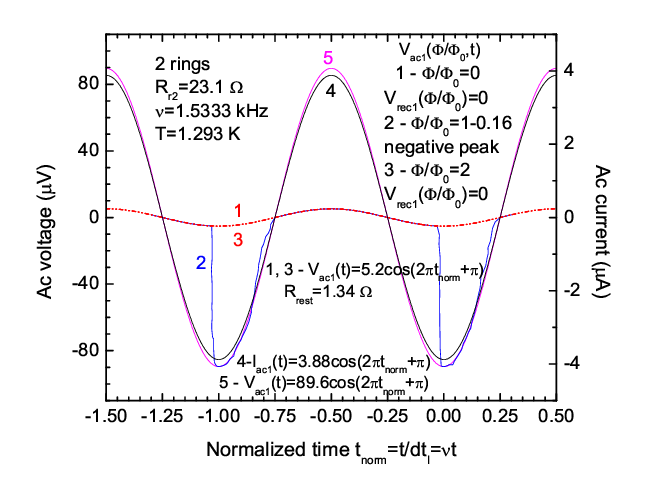}
\caption{\label{f4} (Color online) Lines 1, 2, and 3 are
oscillograms of the ac voltage $V_{ac1}(\Phi/\Phi_{0},t)$, taken
on two rings in series at $\Phi/\Phi_{0}=0$, 1-0.16, and 2,
corresponding to the marked points 1, 2 and 3 (curve
$V_{rec1}(\Phi/\Phi_{0})$, Fig. \ref{f2}), respectively. Lines 1
and 3 are the same. Line 4 is the oscillogram of the applied ac
$I_{ac1}(t)$ (the coordinate axis of the current is located on the
right). Line 5 is the ac voltage $V_{ac1}(t)=I_{ac1}(t)R_{r2}$.}
\end{center}
\end{figure}

Lines 1, 2, and 3 (Fig. \ref{f4}) present the ac voltage
$V_{ac1}(\Phi/\Phi_{0},t)$ at $\Phi/\Phi_{0}=0$, $1-\phi_{l}$ and
2, corresponding to the marked points 1, 2 and 3 (curve
$V_{rec1}(\Phi/\Phi_{0})$, Fig. \ref{f2}), respectively. The
voltage $V_{ac1}(\Phi/\Phi_{0},t)$ is measured on two rings in
series penetrated with the magnetic flux and biased with an
alternating current (without a dc component)
$I_{ac1}(t)=I_{acm1}cos(2\pi t_{norm}+\pi)$ (line 4, Fig.
\ref{f4}). The current scale is shown on the right. Also, for
comparison, line 5 (Fig. \ref{f4}) shows the $\Phi/\Phi_{0}$
independent ac voltage
$V_{ac1}(t)=I_{ac1}(t)R_{r2}=V_{acm1}cos(2\pi t_{norm}+\pi)$
(where $V_{acm1}=I_{acm1}R_{r2}=89.6$ $\mu$V), which is expected
in case the two rings are in a normal state with the total
resistance $R_{r2}$.

The oscillograms of the ac voltage measured at the marked points 1
and 3 are the same. It can be seen that at the marked points 1 and
3, $V_{ac1}(\Phi/\Phi_{0},t)=V_{acm1}cos(2\pi t_{norm}+\pi)$,
where $V_{acm1}= I_{acm1}R_{rest}=5.20$ $\mu$V and $R_{rest}=1.34$
$\Omega$  is a low normal resistance. The low normal resistance
does not depend on $I_{ac1}(t)$ and, possibly from a magnetic
field. This resistance is probably the normal resistance of a
short section of the structure with the lower critical temperature
$T_{c}<1.293$ K. At points 1 and 3 (where $\Phi/\Phi_{0}=0$ and 2)
and for other integer values $\Phi/\Phi_{0}=n$, the rectified
voltage $V_{rec1}(\Phi/\Phi_{0})=0$ (Fig. \ref{f2}).

Note that at the marked point 2, $V_{ac1}(\Phi/\Phi_{0},t)$ occurs
against the background of the low ac voltage equal to
$I_{ac1}(t)R_{rest}$. At the marked point 2 at
$\Phi/\Phi_{0}=1-\phi_{l}$, a negative peak of the voltage -16.74
$\mu$V is observed (Fig. \ref{f2}). It can be seen that the peak
occurs at a current close to the critical current $I_{c}(T,B)$ and
less than $I_{acm1}$.

The average voltage over the period of oscillations of the
harmonic voltage half-wave with amplitude $V_{acm}$  of positive
(or negative) polarity at a sinusoidal alternating current
(without a dc component) is $\pm V_{acm}/\pi$. However, the
voltage occurs when an applied current is close to the switching
critical magnetic-field-dependent current $I_{c}(T,B)$, which
switches the structure from a superconducting state to a resistive
state. Due to the presence of hysteresis $V(I)$ of curves
depending on the direction of the current sweep, the voltage
disappears at low values of the applied current, less than the
retrapping critical current $I_{ret}(T,B)$ for the transition from
the resistive state to the superconducting state. Thus, only about
half of the half-wave contributes to the rectified time-averaged
direct voltage $V_{rec}(\Phi/\Phi_{0})$. Then, if the critical
current is reached only for one half-wave of the current
$I_{ac}(t)$, a rough estimate of the expected maximum voltage
$V_{rec}(\Phi/\Phi_{0})$ at $\Phi/\Phi_{0}=n\pm \phi_{l}$ gives
the value of about $\pm V_{acm}/2\pi$. In particular, a rough
estimate of the expected value $V_{rec1}(\Phi/\Phi_{0})$ for
$\Phi/\Phi_{0}=1- \phi_{l}$ (marked point 2, curve
$V_{rec1}(\Phi/\Phi_{0})$, Fig. \ref{f2}), using the corresponding
voltage oscillogram $V_{ac1}(\Phi/\Phi_{0},t)$ (Fig. \ref{f4}),
gives the value of approximately $-V_{acm1}/2\pi=-14.3 $ $\mu$V. A
more accurate estimate of the average voltage over a period of
time, obtained from the voltage oscillogram
$V_{ac1}(\Phi/\Phi_{0},t)$, gives the value of -14.6 $\mu$V. The
expected value of -14.6 $\mu$V is eighty seven percent of the
measured rectified voltage of -16.7 $\mu$V (marked point 2, curve
$V_{rec1}(\Phi/\Phi_{0})$, Fig. \ref{f2}).

\begin{figure}
\begin{center}
\includegraphics[width=1\linewidth]{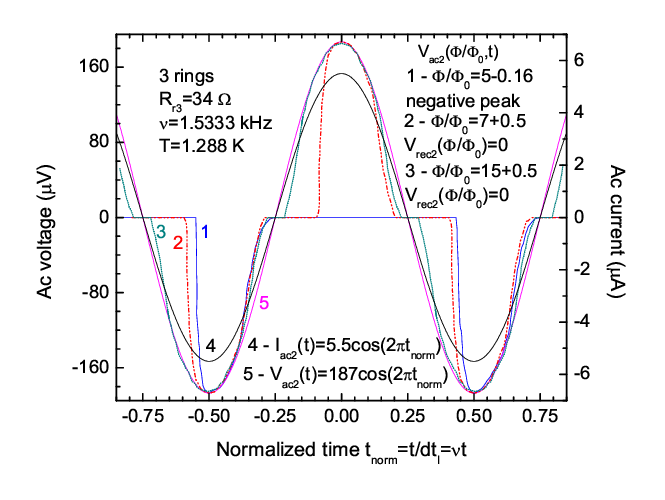}
\caption{\label{f5} (Color online) Lines 1 (solid), 2
(dash-dotted), and 3 (dotted) are oscillograms of the ac voltage
$V_{ac2}(\Phi/\Phi_{0},t)$ recorded on a structure of three rings
in series at $\Phi/\Phi_{0}=5-0.16$, 7+0.5, and 15+0.5,
corresponding to the marked points 1, 2 and 3 (curve
$V_{rec2}(\Phi/\Phi_{0})$, Fig. \ref{f3}), respectively. Line 4 is
the oscillogram of the applied ac $I_{ac2}(t)$ (the coordinate
axis of the current is on the right). Line 5 is the ac voltage
$V_{ac2}(t)=I_{ac2}(t)R_{r3}$.}
\end{center}
\end{figure}

Lines 1 (solid), 2 (dash-dotted), and 3 (dotted) in Fig. \ref{f5}
present the oscillograms of the ac voltage
$V_{ac2}(\Phi/\Phi_{0},t)$ at $\Phi/\Phi_{0}=5-\phi_{l}$, 7+0.5,
and 15+0.5, corresponding to the marked points 1, 2 and 3 (curve
$V_{rec2}(\Phi/\Phi_{0})$, Fig. \ref{f3}), respectively. The
voltage $V_{ac2}(\Phi/\Phi_{0},t)$ is taken on a structure of
three rings in series permeated with a magnetic flux and biased
with an alternating current (without a dc component)
$I_{ac2}(t)=I_{acm2}cos(2\pi t_{norm})$ (line 4 in Fig. \ref{f5}).
The current scale is on the right. In addition, line 5 in Fig.
\ref{f5} shows the ac voltage, independent of $\Phi/\Phi_{0}$,
$V_{ac2}(t)=I_{ac2}(t)R_{r3}=V_{acm2}cos(2\pi t_{norm})$ (here
$V_{acm2}=I_{acm2}R_{r3}=187$ $\mu$V), expected if three rings are
in a normal state with the total resistance $R_{r3}$.

The horizontal sections on the curves $V_{ac2}(\Phi/\Phi_{0},t)$
correspond to the superconducting state. No low normal resistance
is observed. At the marked point 1 (where $\Phi/\Phi_{0}=5-0.16$),
there is a negative peak of the rectified voltage
$V_{rec2}(\Phi/\Phi_{0})$ with the value of -34.8 $\mu$V (Fig.
\ref{f3}). The peak appears at the switching critical current
$I_{c}(T,B)$ close to $I_{acm2}$ and disappears at the retrapping
critical current $I_{ret}(T,B)$ close to zero (Fig. \ref{f5}). At
the marked point 1, a rough and more accurate estimates of the
expected rectified voltage give the values $-V_{acm2}/2\pi=-29.8$
and -34.0 $\mu$V, respectively. At the marked points 2 and 3
(where $\Phi/\Phi_{0}=7+0.5$ and 15+0.5) and also for other
half-integer values $\Phi/\Phi_{0}=n+0.5$, the rectified voltage
$V_{rec2}(\Phi/\Phi_{0})=0$ (Fig. \ref{f3}). It can be seen that
the regions of the superconducting state decrease with increasing
$\Phi/\Phi_{0}$ (Fig. \ref{f5}).

Here, the rectification of an ac voltage has been studied
experimentally in three superconducting circularly-asymmetric
aluminum rings in series permeated with a magnetic flux and biased
with a low-frequency ac (without a dc component) at temperatures
below $T_{c}$.

\subsection{Phase shift of the maxima of the critical currents of different polarity relative to the zero flux in opposite directions along the flux axis (model and experiment)}

Previously it was measured in superconducting
circularly-asymmetric aluminum structures \cite{karpii07,
nikulov07, gurtvoi0910, gurtvoi0603005, burlakov0609345} that the
maxima of the critical currents of different polarity
$I_{c+}(\Phi/\Phi_{0}+\phi_{l})$ and
$I_{c-}(\Phi/\Phi_{0}-\phi_{l})$ are shifted by $\phi_{l}$
relative to the zero flux in opposite directions along the
normalized flux axis. The mutual shift of the maxima of the
critical currents relative to each other $2\phi_{l}$ reaches the
value of 0.5, corresponding to half of the oscillation period.

In this section, we sketch a model for the appearance of the shift
$\phi_{l}=\Delta \Phi/\Phi_{0}$ of the maxima of the critical
currents of different polarity $I_{c+}(\Phi/\Phi_{0}+\phi_{l})$
and $I_{c-}(\Phi/\Phi_{0}-\phi_{l})$) relative to the zero flux in
opposite directions along the flux axis in a superconducting
thin-film circularly-asymmetric aluminum ring permeated with a
magnetic flux at $T$ close to $T_{c}$. Oscillations of the
critical current in a superconducting ring of small radius $r$,
penetrated by a magnetic flux, are determined by the requirement
of quantizing the superconducting fluxoid \cite{tinkham}. When the
ring has $r$ greater than $\xi(T)$ and the ring is
circularly-asymmetric, then the ring cannot be regarded as a
homogeneous superconducting structure. In this case, to calculate
the oscillations of the critical current as a function of the
flux, a more complex problem must be solved. In \cite{karpii07,
nikulov07, gurtvoi0910, gurtvoi0603005, burlakov0609345}, the
critical currents of different polarity as functions of the flux
were measured in superconducting circularly-asymmetric aluminum
structures in a stationary situation with the zero voltage on the
structures, since the applied current did not practically exceed
the critical values. We believe that the asymmetric ring can work
as an asymmetric dc micro-SQUID \cite{barone, clarke}. Therefore,
we apply the theory of the stationary superconducting states of
the asymmetric dc SQUID \cite{clarke} to create our model. Our
thin-film interferometer has a thickness $d$ much smaller than the
temperature-dependent London penetration depth of the magnetic
field ($d<<\lambda(T)$) \cite{schmidt}. The interferometer
consists of a narrow and wide semirings (arms) of approximately
the same length $l_{r}=\pi r$, with widths $w_{n}$ and $w_{w}$,
respectively.

We believe that the total temperature-dependent inductances of the
wide and narrow arms are given by the expressions
$L_{1}(T)=L_{m1}(T)+L_{k1}(T)$ and $L_{2}(T)=L_{m2}(T)+L_{k2}(T)$,
respectively. Here $L_{k1}(T)$ and $L_{k2}(T)$ are the kinetic
inductances of the wide and narrow arms, respectively, $L_{m1}(T)$
and $L_{m2}(T)$ are the magnetic inductances of the wide and
narrow arms, respectively. For $d<<\lambda(T)$, the magnetic
inductance $L_{m}(T) \approx 0.002L_{k}(T)$, therefore, in what
follows we set $L_{1}(T)=L_{k1}(T)$ and $L_{2}(T)=L_{k2}(T)$.

The kinetic inductances \cite{schmidt} of both wide and narrow
arms of an asymmetric interferometer are given by the expressions
in the international system of units (SI)
$L_{k1}(T)=\mu_{0}\lambda_{1}^{2}(T)l_{r}/w_{w}d$ and
$L_{k2}(T)=\mu_{0}\lambda_{2}^{2}(T)l_{r}/w_{n}d$, respectively
(where $\mu_{0}=4\pi\times10^{-7}$ H/m, $\lambda_{1}(T)$ and
$\lambda_{2}(T)$ are temperature-dependent London depths for wide
and narrow arms, respectively).

We introduced the dimensionless inductances $l_{1}(T)=2\pi
L_{1}(T)I_{c1}(T)/\Phi_{0}$ and $l_{2}(T)=2\pi
L_{2}(T)I_{c2}(T)/\Phi_{0}$, where $I_{c1}(T)$ and $I_{c2}(T)$ are
the critical currents of wide and narrow arms of the
interferometer, respectively. Here $L_{1}(T)=L_{k1}(T)$ and
$L_{2}(T)=L_{k2}(T)$. The total dimensionless inductance of the
interferometer is $l(T)=l_{1}(T)+l_{2}(T)$. In the limit
$l(T)>>1$, the difference
$L_{1}(T)I_{c1}(T)/\Phi_{0}-L_{2}(T)I_{c2}(T)/\Phi_{0}$ is equal
to the shift $\phi_{l}$ of the maxima in the critical currents of
the interferometer $I_{c+}(\Phi/\Phi_{0}+\phi_{l})$ and
$I_{c-}(\Phi/\Phi_{0}-\phi_{l})$ of different polarity with
respect to the zero flux in opposite directions along the
normalized flux axis \cite{clarke}. The phase shift of the maxima
of the critical currents of the interferometer is equal to the
$dl(T)=l_{1}(T)-l_{2}(T)$ and corresponds to the shift
$\phi_{l}(T)=\Delta \Phi/\Phi_{0}$ (which determined in fractions
of the oscillation period $\Phi_{0}$), so that the equality
$dl(T)=2 \pi \phi_{l}(T)$ holds.

If both arms of the asymmetric interferometer are superconducting
and have the same length and thickness, the difference $dl(T)$ can
be easily calculated.

Taking into account the expressions for the temperature-dependent
kinetic inductances $L_{k}(T)$ and the GL depairing critical
current densities in SI
$j_{c}(T)=\Phi_{0}/(\sqrt{27}\pi\mu_{0}\lambda^{2}(T)\xi(T))$ and
the expressions for the critical currents
$I_{c1}(T)=j_{c1}(T)w_{w}d$ and $I_{c2}(T)=j_{c2}(T)w_{n}d$ of the
wide and narrow arms of the interferometer,
$dl(T)=l_{1}(T)-l_{2}(T)=(2l_{r}/\sqrt{27})(1/\xi_{1}(T)-1/\xi_{2}(T))$
is obtained. Note that $dl(T)$ does not depend on the width and
thickness of the ring and is equal to zero at
$\xi_{1}(T)=\xi_{2}(T)$, that is, if the critical current
densities $j_{c1}(T)$ and $j_{c2}(T)$ are the same in the wide and
narrow arms of the interferometer.

Thus, the difference $dl(T)$ is not zero only if the densities
$j_{c1}(T)$ and $j_{c2}(T)$ are different. This may be the case
when the critical temperatures $T_{cw}$ and $T_{cn}$ are not the
same for the wide and narrow arms, respectively.

Note that, in this model, we neglect the dependences of the
critical current densities on the magnetic field since in the
experimental range of fields \cite{karpii07, nikulov07} these
current densities do not practically change. This is due to the
fact that the upper experimental field boundary is well below the
maximum critical fields at which the superconducting order
parameter is completely suppressed in the narrow and wide arms of
the interferometer.

Nonlocal and local negative voltages found in
\cite{kuznphysica20} can be explained  by the assumption that  the
critical temperature of a wide superconducting
quasi-one-dimensional aluminum structure is higher than the
critical temperature of a narrow structure with the same thickness
of the structures. To check the condition $T_{cw}>T_{cn}$, and
hence the inequality $j_{c1}(T)>j_{c2}(T)$, the critical
temperatures and critical currents are measured depending on $T$
in the zero magnetic field in wide and narrow superconducting
aluminum structures with dimensions close to the typical
dimensions of the wide and narrow arms of the interferometer.
Indeed, these measurements showed that the critical temperature of
a superconducting thin-film quasi-one-dimensional aluminum wire is
smaller, the smaller the wire width (the results will be presented
elsewhere).

Thus, in our model, the phase shift $dl(T)=2 \pi \phi_{l}(T)$
appears due to the difference between the critical temperatures
$T_{cw}$ and $T_{cn}$. We apply this model to the experimental
shift $\phi_{l}(T)$ of the maxima of the critical currents of
different polarity $I_{c+}(\Phi/\Phi_{0}+\phi_{l})$ and
$I_{c-}(\Phi/\Phi_{0}-\phi_{l})$) relative to the zero flux in
opposite directions along the flux axis in six superconducting
circularly-asymmetric aluminum structures measured in
\cite{karpii07, nikulov07, gurtvoi0910, gurtvoi0603005,
burlakov0609345}.

At $T<T_{cn}$, both arms of the circularly-asymmetric
interferometer are superconducting and have effective critical
temperatures $T_{cf1}$ and $T_{cf2}$ (taking into account the
proximity effect) for the wide and narrow arms, respectively. At
$T<T_{cn}$, the phase shift of the maxima of the critical currents
of different polarity with respect to the zero flux in opposite
directions along the flux axis is determined by the equation
$dl(T<T_{cn})=l_{1}(T<T_{cn})-l_{2}(T<T_{cn})$, where the
dimensionless inductances of the wide and narrow arms
$l_{1}(T<T_{cn})$ and $l_{2}(T<T_{cn})$ are given by the
expressions $l_{1}(T<T_{cn})=2\pi
L_{1}I_{c1}/\Phi_{0}=(2l_{r}/\sqrt{27})(1/\xi_{1}(T)=(2\pi
r/(\sqrt{27}\xi(0)))\sqrt{1-T/T_{cf1}}$ and $l_{2}(T<T_{cn})=2\pi
L_{2}I_{c2}/\Phi_{0}=(2l_{r}/\sqrt{27})(1/\xi_{2}(T)=(2\pi
r/(\sqrt{27}\xi(0)))\sqrt{1-T/T_{cf2}}$, respectively.

We believe that at temperatures $T_{cn}<T<T_{cw}$, the wide arm of
the interferometer is superconducting, while the narrow arm is a
hybrid Josephson S-s-S structure, in the center of which the
superconducting order parameter is weakened. In the intervals
$T_{cn}<T<T_{cw}$ and $T<T_{cn}$, dimensionless inductances of the
wide arm are given by the same expression
$l_{1}(T_{cn}<T<T_{cw})=l_{1}(T<T_{cn})=(2\pi
r/(\sqrt{27}\xi(0)))\sqrt{1-T/T_{cf1}}$. At $T_{cn}<T<T_{cw}$, the
dimensionless inductance of the narrow arm is determined by
another expression $l_{2}(T_{cn}<T<T_{cw})=2\pi
L_{2}(T)I_{c2}(T)/\Phi_{0}$, where
$L_{2}(T)=\mu_{0}\lambda_{2}^{2}(T)\pi r/w_{n}d$ is the inductance
of the narrow arm, $I_{c2}(T)$ is the critical current of the
narrow arm. We believe that $I_{c2}(T)$ coincides with the
critical current of the Josephson structure $I_{J}(T)$. Near
$T_{c}$, $I_{J}(T)=\pi \Delta^{2}(T)/(4ekTR_{J})$, where
$\Delta(T)=1.74\Delta(0)(1-T/T_{c})^{1/2}$ is the
temperature-dependent energy superconducting gap in the zero
magnetic field, $\Delta(0)=1.764kT_{c}$ is the gap at $T=0$,
$R_{J}$ is the Josephson resistance \cite{likharev}. We can write
$I_{J}(T)=I_{J}(0)(1-T/T{c})$, where $I_{J}(0)$ is the Josephson
critical current at $T=0$ K. Thus, near $T_{c}$, the currents
$I_{J}(T)$ and $I_{c2}(T)$ are linear functions of $T$.

The linear temperature dependence of the critical current of the
narrow arm of the interferometer $I_{c2}(T)$ is confirmed by the
linear dependence of the experimental switching and retrapping
critical currents in a narrow aluminum structure with dimensions
close to the dimensions of the narrow arm of the interferometer,
in the temperature range when the narrow structure taken as a
whole is a hybrid S-s-S structure. In addition, in this interval,
the switching and retrapping currents are the same. Measurement
results will be reported elsewhere.

Then the dimensionless inductance of the narrow arm is given by
the expression $l_{2}(T_{cn}<T<T_{cw})=K\mu_{0}\lambda_{2}^{2}(T)
rT_{cf2}(1-T/T_{cf2})/(\Phi_{0}\rho_{n}l_{J})$, where the
Josephson resistance is taken into account in the form
$R_{J}=\rho_{n} l_{J}/(w_{n}d)$, where $l_{J}$ is the effective
Josephson length, $\rho_{n}$ is the resistivity, $r$ is the
average radius of the circularly-asymmetric interferometer,
$K=(\pi^{3}/2)(1.74*1.764)^{2}(k/e)=1.259\times10^{-2}$ V/K.

We believe that at temperatures $T_{cn}<T<T_{cw}$, the Josephson
S-s-S junction forms naturally in the middle of the narrow arm due
to the proximity effect and equilibrium fluctuations and has a
length $l_{J}$ not exceeding $2\xi(T)$. Thus, near the critical
temperature, it is expected that the current-phase relationship at
the Josephson junction will be a single-valued function, with a
shape slightly deviating from the sinusoidal shape
\cite{likharev}.

Recall that at temperatures $T<T_{cw}$ and $T<T_{cn}$ for the wide
and narrow arms of the interferometer, respectively, the critical
currents of both arms $I_{c1}(T)$ and $I_{c2}(T)$ are determined
by the depairing critical current densities $j_{c1}(T)$ and
$j_{c2}(T)$ in both arms. However, we believe that at temperatures
close to critical temperatures, both arms are similar to
self-generating long Josephson S-s-S contacts. It is expected that
the current-phase relation in the arms will be a single-valued
function with a shape that does not deviate greatly from the
sinusoidal shape, since the effective length of the arms does not
exceed the value $3\xi(T)$ \cite{likharev}.

Using the expression for
$\lambda_{2}(T)=\lambda(0)(1-T/T_{cf2})^{-1/2}$, we obtain the
expression $l_{2}(T_{cn}<T<T_{cw})=K\mu_{0}\lambda(0)^{2}
rT_{cf2}/(\Phi_{0}\rho_{n}l_{J})$, which does not explicitly
depend on temperature, width, thickness. In this expression, two
adjustable parameters are the effective critical temperature of
the narrow arm $T_{cf2}$ and the Josephson length $l_{J}$.

Further, we use this model of the temperature-dependent phase
shift $dl(T)=2 \pi \phi_{l}(T)$ of the maxima of the critical
currents of different polarity in superconducting
circularly-asymmetric aluminum structures to describe the
temperature-dependent experimental shift $\phi_{l}(T)$ of the
maxima of the critical currents of different polarity
$I_{c+}(\Phi/\Phi_{0}+\phi_{l}(T))$ and
$I_{c-}(\Phi/\Phi_{0}-\phi_{l}(T))$) relative to the zero flux in
opposite directions along the flux axis.

We used the results presented earlier in \cite{nikulov07,
gurtvoi0910, gurtvoi0603005, burlakov0609345}. Authors of
\cite{nikulov07, gurtvoi0910, gurtvoi0603005, burlakov0609345}
measured critical currents for opposite directions of the applied
current $I_{c+}(\Phi/\Phi_{0}+\phi_{l}(T))$ and
$I_{c-}(\Phi/\Phi_{0}-\phi_{l}(T))$  as functions of the
normalized magnetic flux in superconducting single
circularly-asymmetric aluminum rings and such rings in series with
different parameters, including different values of asymmetry, at
different temperatures.

Since the main parameters of superconducting structures
\cite{nikulov07, gurtvoi0910, gurtvoi0603005, burlakov0609345} are
not given exactly, these parameters will be estimated. The mean
free path of electrons $l_{el}=15.64-18.4$ nm is obtained from the
expression \cite{ashcroft} $l_{el}=l_{ph}(R_{300}/R_{4.2}-1)$,
where $l_{ph}=9.2$ nm is the effective electron path length for
scattering by phonons for aluminum at $T=300$ K,
$R_{300}/R_{4.2}\approx 2.7-3.0$ is the resistance ratio for
thin-film circularly-asymmetric aluminum structures with the film
thickness of 10-60 nm \cite{nikulov07, gurtvoi0910,
gurtvoi0603005, burlakov0609345}. The resistivity of the structure
$\rho_{n}=3.26-2.77 \times 10^{-8}$ $\Omega$\,m was obtained from
the refined theoretical expression for aluminum \cite{gershenson}
$\rho_{n}l_{el}=5.1\times 10^{-16}$ $\Omega$\,m$^{2}$. Since
$l_{el}<<\xi_{0}=1.6$ $\mu$m, then near $T_{c}$, the
Ginzburg-Landau coherence length
$\xi(T)=\xi(0)(1-T/T_{c})^{-1/2}$, where
$\xi(0)=0.85(l_{el}\xi_{0})^{1/2}$. The temperature-dependent
penetration depth of the magnetic field is
$\lambda_{GL}(T)=\lambda(0)(1-T/T_{c})^{-1/2}$, where
$\lambda(0)=0.615\lambda_{L}(\xi_{0}/l_{el})^{1/2}$.

We used the following parameters: $l_{el}=15.6$ nm, $\xi(0)=0.134$
$\mu$m, $\lambda(0)=0.100$ $\mu$m, $\rho_{n}=3.26\times 10^{-8}$
$\Omega$\,m for aluminum structures \cite{nikulov07, gurtvoi0910}
with the thickness of $d=20-10$ nm.

The following parameters: $l_{el}=18.4$ nm, $\xi(0)=0.146$ $\mu$m,
$\lambda(0)=0.092$ $\mu$m, $\rho_{n}=2.77\times 10^{-8}$
$\Omega$\,m are taken for aluminum structures \cite{nikulov07,
gurtvoi0910, gurtvoi0603005, burlakov0609345} with the thickness
of $d=35-60$ nm.

We took the experimental values of the temperature-dependent shift
$\phi_{l}(T)$ of the maxima of the critical currents of different
polarity relative to the zero flux along the flux axis directly
from the curves of critical currents of different polarity
$I_{c+}(\Phi/\Phi_{0}+\phi_{l}(T))$ and
$I_{c-}(\Phi/\Phi_{0}-\phi_{l}(T))$ measured at different
temperatures in superconducting circularly-asymmetric aluminum
structures: "ring 1" \cite{nikulov07}, "ring 2"
\cite{gurtvoi0603005, burlakov0609345}, "ring 3"
\cite{gurtvoi0603005, burlakov0609345}, "18 rings"
\cite{nikulov07}, "ring" \cite{gurtvoi0910} and "another ring"
with similar parameters \cite{burlakov0609345}, "110 rings"
\cite{gurtvoi0910}. The corresponding experimental phase shift was
determined by the expression $dl_{exp}(T)=2 \pi \phi_{l}(T)$.

The ring "ring 1" is a single circularly-asymmetric aluminum ring
with a resistance in the normal state $R_{n}=3.3$ $\Omega$, with
an average radius $r=2.1$ $\mu$m, with a current wire width
$w_{I}=0.7$ $\mu$m, thickness $d_{f}=60$ nm, widths of wide and
narrow semirings $w_{w}=0.4$, and $w_{n}=0.2$ $\mu$m, respectively
\cite{nikulov07}. The critical temperatures determined from the
three levels of the N-S transition $R(T)/R_{n}=0$, 0.5, 0.96 are
$T_{cl}(0)=1.237$, $T_{c}(0.5)=1.247$, and $T_{ch}(0.96)=1.279$ K,
respectively. The length of the N-S transition is
$dT=T_{ch}(0.96)-T_{cl}(0)=0.042$ K \cite{nikulov07}.

The ring "ring 2" is a single circularly-asymmetric aluminum ring
with dimensions: $r=2.1$ $\mu$m, $d_{f}=40$ nm, $w_{w}=0.25$, and
$w_{n}=0.2$ $\mu$m and the critical temperature $T_{c}(0.5)=1.275$
K \cite{gurtvoi0603005, burlakov0609345}.

The ring "ring 3" is a single circularly-asymmetric aluminum ring
with dimensions: $r=2.1$ $\mu$m, $d_{f}=35$ nm, $w_{w}=0.3$, and
$w_{n}=0.2$ $\mu$m and the critical temperature $T_{c}(0.5)=1.288$
K \cite{gurtvoi0603005, burlakov0609345}.

The structure "18 rings" with a resistance in the normal state
$R_{n}=92$ $\Omega$ consists of 18 identical circularly-asymmetric
aluminum rings in series connected by a current wire $w_{I}=0.4$
$\mu$m wide. Each ring has dimensions: $r=2.1$ $\mu$m, $d_{f}=20$
nm, $w_{w}=0.4$, and $w_{n}=0.2$ $\mu$m. The critical temperature
is $T_{c}(0.5)=1.270$ K \cite{nikulov07}.

The structure "ring" is a single circularly-asymmetric aluminum
ring with dimensions: $r=2.1$ $\mu$m, $d_{f}=35$ nm, $w_{w}=0.3$,
$w_{n}=0.2$ $\mu$m and the critical temperature $T_{c}(0.5)=1.294$
K \cite{gurtvoi0910}.

The structure "110 rings" with a resistance in the normal state
$R_{n}=960$ $\Omega$ (one ring corresponds to the resistance
$R_{n1}=960/110=8.72$ $\Omega$) consists of 110 identical
circularly-asymmetric aluminum rings in series. Each ring has
dimensions: $r=1.1$ $\mu$m, $d_{f}=10$ nm, $w_{w}=0.4$, and
$w_{n}=0.2$ $\mu$m. The critical temperature is $T_{c}(0.5)=1.345$
K \cite{gurtvoi0910}.

Note that the above values of the thicknesses $d_{f}$ of six
structures are given taking into account the correction after
using the thickness as an adjustable parameter when comparing the
experimental $L_{exp}(T)$ and theoretical $L_{1}(T)+L_{ 2}(T)$
temperature dependences of the total inductance of one ring
included in the structure. See section 2.3.

\begin{figure}
\begin{center}
\includegraphics[width=1\linewidth]{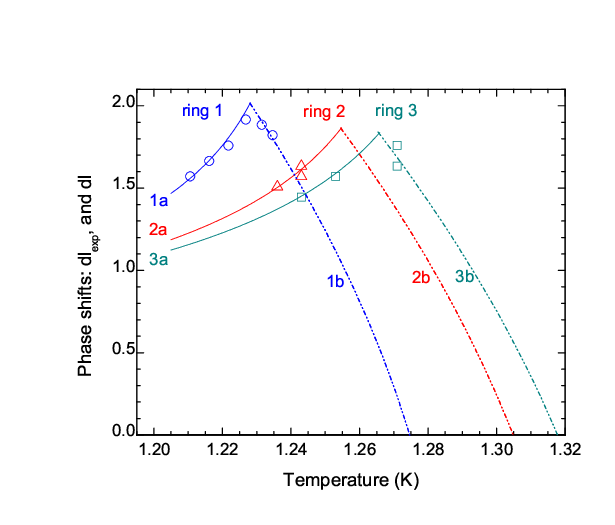}
\caption{\label{f6} (Color online) Circles, triangles and squares
are the measured temperature-dependent phase shifts $dl_{exp}(T)=2
\pi \phi_{l}(T)$ of the maxima of the critical currents of
different polarity with respect to the zero flux in opposite
directions along the flux axis in three single
circularly-asymmetric aluminum rings: "ring 1", "ring 2", and
"ring 3", respectively. Three pairs of lines 1a-1b, 2a-2b, and
3a-3b, including ascending (a) and descending (b) branches are
fits of the phase shifts (circles, triangles and squares) using
the corresponding theoretical functions $dl(T)$ for three rings:
"ring 1", "ring 2", and "ring 3", respectively.}
\end{center}
\end{figure}

Figure \ref{f6} shows the experimental phase shifts $dl_{exp}(T)=2
\pi \phi_{l}(T)$ of the maxima of the critical currents of
different polarity with respect to the zero flux in opposite
directions along the flux axis (circles, triangles, and squares)
and the corresponding adjustable theoretical curves $dl(T)$
(1a-1b, 2a-2b, and 3a-3b) as functions of $T$ for three single
superconducting circularly-asymmetric aluminum rings: "ring 1",
"ring 2", and "ring 3", respectively.

In Fig. \ref{f6}, we plotted the experimental phase shifts
$dl_{exp}(T)=2 \pi \phi_{l}(T)$ of the maxima of the critical
currents of different polarity with respect to the zero flux in
opposite directions along the flux axis at temperatures $T=1.211$,
1.216, 1.222, 1.227, 1.232, and 1.235 K on the "ring 1" (circles),
at $T=1.236$, 1.243, and 1.243 K on "ring 2" (triangles), at
$T=1.243$, 1.253, and 1.271 K on the "ring 3" (squares).
Experimental data are approximated by three pairs of adjustable
curves 1a-1b, 2a-2b, and 3a-3b (Fig. \ref{f6}).

\begin{figure}
\begin{center}
\includegraphics[width=1\linewidth]{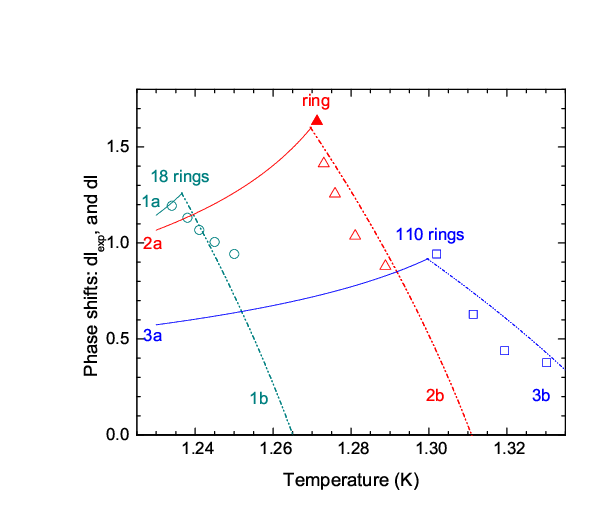}
\caption{\label{f7} (Color online) Circles, triangles and squares
are the experimental temperature-dependent phase shifts
$dl_{exp}(T)=2 \pi \phi_{l}(T)$ of the maxima of the critical
currents of different polarity with respect to the zero flux in
opposite directions along the flux axis for three
circularly-asymmetric structures: "18 rings", "ring", and "110
rings", respectively. Closed triangle is the measured phase shift
at $T=1.271$ K for "another ring" with the same parameters and
close critical temperature $T_{c}(0.5)=1.288$ K
\cite{burlakov0609345}. Lines 1a-1b, 2a-2b, and 3a-3b, each of
which consists of increasing (a) and decreasing (b) branches are
the phase shift fits (circles, triangles and squares) using the
appropriate $dl(T)$ functions for three structures: "18 rings",
"ring", and "110 rings", respectively.}
\end{center}
\end{figure}

Figure \ref{f7} shows the experimental temperature-dependent phase
shifts of the maxima of the critical currents of different
polarity with respect to the zero flux in opposite directions
along the flux axis (circles, triangles and squares) and the
corresponding adjustable theoretical $dl(T)$ curves (1a-1b, 2a-2b,
and 3a-3b) for three superconducting circularly-asymmetric
aluminum structures: "18 rings", "ring", and "110 rings",
respectively.

In Fig. \ref{f7}, we presented the measured phase shifts
$dl_{exp}(T)=2 \pi \phi_{l}(T)$ of the maxima of the critical
currents of different polarity with respect to the zero flux in
opposite directions along the flux axis at $T=1.234$, 1.238,
1.241, 1.245 K on the structure "18 rings" (circles), at
$T=1.273$, 1.276, 1.281, and 1.289 K on the structure "ring"
(triangles) and at $T=1.271$ K (closed triangle) on "another ring"
with the same dimensions and close $T_{c}(0.5)=1.288$ K
\cite{burlakov0609345}, at $T=1.302$, 1.311, 1.320, and 1.330 K
for the structure "110 rings" (squares). Experimental data are
approximated by three pairs of adjustable curves 1a-1b, 2a-2b, and
3a-3b (Fig. \ref{f7}).

We found that the phase shift $dl(T)$ of the maxima of the
critical currents is a non-monotonic function of $T$ (Figs.
\ref{f6}, \ref{f7}).

Recall that at $T<T_{cn}$, the phase shift of the maxima of the
critical currents of different polarity with respect to the zero
flux in opposite directions along the flux axis is determined by
the difference of  the dimensionless inductances of the wide and
narrow arms $dl(T<T_{cn})=l_{1}(T<T_{cn})-l_{2}(T<T_{cn})$. The
phase shifts of the maxima of the critical currents of different
polarity with respect to the zero flux in opposite directions
along the flux axis (curves 1a-3a of Fig. \ref{f6} and curves
1a-3a of Fig. \ref{f7}) are given by the equation
\begin{equation}
\label{e1} dl(T<T_{cn})=\frac{2\pi
r(\sqrt{1-T/T_{cf1}}-\sqrt{1-T/T_{cf2}})}{\sqrt{27}\xi(0)},
\end{equation}
where the values of the physical parameters: $r$, $\xi(0)$ and the
adjustable effective (taken into account the proximity effect)
critical temperatures of the wide and narrow semirings $T_{cf1}$
and $T_{cf2}$, respectively, are shown in the Table \ref{t1}.

Note that at $T_{cn}<T<T_{cw}$, the phase shift of the maxima of
the critical currents of different polarity with respect to the
zero flux in opposite directions along the flux axis is given by
the difference
$dl(T_{cn}<T<T_{cw})=l_{1}(T_{cn}<T<T_{cw})-l_{2}(T_{cn}<T<T_{cw})$.
The phase shifts of the maxima of the critical current (curves
1b-3b of Fig. \ref{f6} and curves 1b-3b of Fig. \ref{f7}) are
described by the equation

\begin{equation}
\label{e2} dl(T_{cn}<T<T_{cw})=\frac{2\pi r
\sqrt{1-T/T_{cf1}}}{\sqrt{27}\xi(0)}- \frac{K
\mu_{0}\lambda(0)^{2} rT_{cf2}}{\Phi_{0}\rho_{n} l_{J}}
\end{equation}
here the values of the physical parameters: $r$, $\xi(0)$,
$\lambda(0)$, $\rho_{n}$\, the adjustable critical temperatures
$T_{cf1}$ and $T_{cf2}$ are the same as for equation (\ref{e1}),
and the adjustable Josephson length $l_{J}$ are presented in the
Table \ref{t1}.

\begin{table*}
\caption{\label{t1} $r$ is the average radius of the ring from a
structure, $\xi(0)$ is the Ginzburg-Landau coherence length at
$T=0$, $\lambda(0)$ is the penetration depth of the magnetic field
at $T=0$ K, $\rho_{n}$ is the structure resistivity, $T_{c}(0.5)$
is the critical temperature determined on the mean level of the
N-S transition of the structure, $T_{cf1}$ and $T_{cf2}$ are the
adjustable effective critical temperatures of the wide and narrow
semirings, respectively, $T_{cn}$ is the crossover temperature
corresponding to the maximum value of $dl(T)$, $l_{J}$ is the
adjustable Josephson length.}
\centering %
\begin{tabular}{cccccccccc}
\hline
structure & $r$, $\mu$m & $\xi(0)$, $\mu$m & $\lambda(0)$, $\mu$m & $\rho_{n}$, $\Omega$\,$\mu$m & $T_{c}(0.5)$, K & $T_{cf1}$, K & $T_{cf2}$, K & $T_{cn}$, K & $l_{J}$, $\mu$m \\
\hline
ring 1 (1a-1b, Fig. \ref{f6}) & 2.1 & 0.146 & 0.092 & 0.0277 & 1.247 & 1.287 & 1.240 & 1.229 & 3.55 \\
ring 2 (2a-2b, Fig. \ref{f6}) & 2.1 & 0.146 & 0.092 & 0.0277 & 1.275 & 1.325 & 1.274 & 1.255 & 2.90 \\
ring 3 (3a-3b, Fig. \ref{f6}) & 2.1 & 0.146 & 0.092 & 0.0277 & 1.288 & 1.341 & 1.288 & 1.268 & 2.75 \\
\hline
18 rings (1a-1b, Fig. \ref{f7}) & 2.1 & 0.134 & 0.100 & 0.0326 & 1.270 & 1.288 & 1.259 & 1.237 & 2.43 \\
ring (2a-2b, Fig. \ref{f7}) & 2.1 & 0.146 & 0.092 & 0.0277 & 1.294 & 1.331 & 1.289 & 1.272 & 2.95 \\
110 rings (3a-3b, Fig. \ref{f7}) & 1.1 & 0.134 & 0.100 & 0.0326 & 1.345 & 1.385 & 1.332 & 1.303 & 2.20 \\
\hline
\end{tabular}
\end{table*}

Table \ref{t1} shows the physical and fitting parameters used in
equations (\ref{e1}) and (\ref{e2}) to calculate the temperature
dependence of the phase shift $dl(T)$ of the maxima of critical
currents of different polarity relative to zero magnetic flux in
different directions along the flux axis in the six structures
under study. These parameters are an average radius $r$ of the
ring from a structure, the Ginzburg-Landau coherence length
$\xi(0)$ at $T=0$, the penetration depth $\lambda(0)$ of the
magnetic field at $T=0$ K, the structure resistivity $\rho_{n}$,
the critical temperature $T_{c}(0.5)$ determined on the mean level
of the N-S transition of the structure, the adjustable effective
critical temperatures $T_{cf1}$ and $T_{cf2}$ of the wide and
narrow semirings, respectively, the crossover temperature $T_{cn}$
corresponding to the maximum value of $dl(T)$, the adjustable
Josephson length $l_{J}$.

We assume that for all circularly-asymmetric aluminum structures,
the adjustable effective critical temperature (taking into account
the proximity effect) of the wide semiring $T_{cf1}$ corresponds
to the top of the resistive N-S transition. The Table \ref{t1}
shows that in most cases, the adjustable effective critical
temperature (taking into account the proximity effect) of the
narrow semiring $T_{cf2}$ is below the critical temperature
$T_{c}(0.5)$. We assume that $T_{cf2}$ is close to the critical
temperature $T_{cl}(0)$, corresponding to the bottom of the
resistive N-S transition (Figs. \ref{f6}, \ref{f7}).

The assumed true critical temperature of a narrow semiring
(without taking into account the influence of the proximity
effect) $T_{cn}$, determined by the intersection of the adjustable
curves $dl(T<T_{cn})$ and $dl(T_{cn}<T<T_{cw})$ and corresponding
to the maximum value of $dl(T)$, can be significantly lower than
the critical temperature $T_{cl}(0)$. We believe that the
difference $T_{cf1}-T_{cn}(0)$ is close to the difference between
the true temperatures (without taking into account the influence
of the proximity effect) of the wide and narrow semirings (Figs.
\ref{f6}, \ref{f7}). This difference can reach 87 mK.

As expected, the data from the Table \ref{t1} confirms that the
adjustable Josephson length $l_{J} \approx 2\xi(T)$.

Moreover, for each structure, after plotting an adjustable
theoretical temperature dependence of the phase shift equal to the
difference between the dimensionless inductances of the wide and
narrow semirings $dl(T)=l_{1}(T)-l_{2}(T)$, we calculated the
temperature dependence of the total dimensionless inductance of an
individual ring from the structure $l(T)=l_{1}(T)+l_{2}(T)$. The
functions $l(T)$ are not shown in the figures. However, the ranges
of $l(T)$ values in certain temperature $T$ ranges are given in
the Table \ref{t2} for each structure. We found that the
calculated total dimensionless temperature-dependent inductance of
an individual ring from the structures under study is
$l(T)=l_{1}(T)+l_{2}(T)=8-4$ in the temperature ranges under
study. Thus, the case of a sufficiently large total inductance of
the ring $l>>1$ is almost realized.

\begin{table}
\caption{\label{t2} $d_{f}$ is the adjustable thickness of the
structure, $R_{J}$ is the Josephson resistance, $\pi r/l_{J}$ is
the ratio of the semiring length to the Josephson length, $T$ is
the temperature interval, $l(T)=l_{1}(T)+l_{2}(T)$ - the range of
changes in the calculated total dimensionless inductance of a
separate ring from the structure.}
\centering %
\begin{tabular}{cccccccccc}
\hline
structure& $d_{f}$, nm & $R_{J}$, $\Omega$ & $\pi r/l_{J}$ & $T$, K & $ l(T)$ \\
\hline
ring 1& 60 & 8.20 & 1.86 & 1.210-1.235 & 7.0-5.2 \\
ring 2& 40 & 10.0 & 2.28 & 1.235-1.245 & 7.6-6.9 \\
ring 3& 35 & 10.9 & 2.40 & 1.245-1.270 & 7.8-6.3 \\
\hline
18 rings& 20 & 19.8 & 2.72 & 1.234-1.250 & 6.6-5.8 \\
ring& 35 & 11.7 & 2.24 & 1.270-1.290 & 5.9-5.2 \\
110rings& 10 & 35.9 & 1.57 & 1.300-1.330 & 4.0-3.5 \\
\hline
\end{tabular}
\end{table}
Table \ref{t2} shows the values of the adjustable thickness
$d_{f}$ of the structures obtained after comparing the
experimental $L_{exp}(T)$ and theoretical $L_{1}(T)+L_{ 2}(T)$
temperature dependences of the total inductance of one ring
included in the structure. See section 2.3.

In addition, for each structure, the Table \ref{t2} shows the
values of the Josephson resistance $R_{J}$ and the ratio $\pi
r/l_{J}$ of the semiring length to the Josephson length, which
were obtained using the parameters of Tables \ref{t1} and
\ref{t2}.

The values of the temperature interval $T$ and the corresponding
values of the range of variation of the calculated total
dimensionless inductance $l(T)=l_{1}(T)+l_{2}(T)$ of an individual
ring of the structure are placed in the last two columns of the
Table \ref{t2} for each structure.

\subsection{Temperature dependences of calculated inductances $L_{1}(T)$ and $L_{2}(T)$  of  wide and narrow semirings and experimental inductance of the ring $L_{exp}(T)$}

\begin{figure}
\begin{center}
\includegraphics[width=1\linewidth]{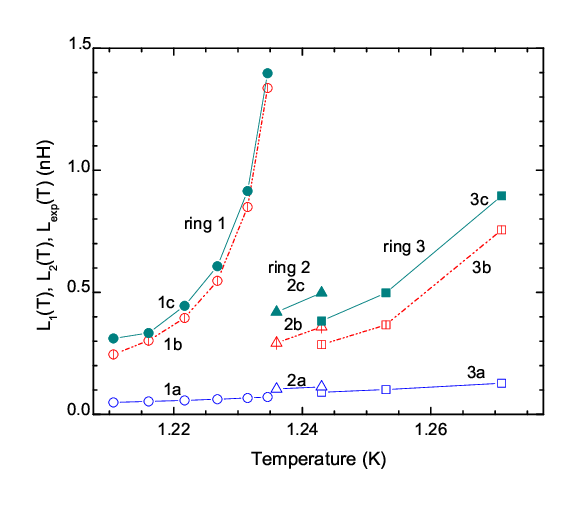}
\caption{\label{f8} (Color online) Curve 1a (open circles), curve
1b (open circles with a vertical bar), curve 1c (closed circles)
are the theoretical inductances of wide, narrow semirings
$L_{1}(T)$, $L_{2}(T)$ and experimental inductance of one ring
$L_{exp}(T)$ as functions of $T$ in the range 1.211-1.235 K,
respectively, for the structure "ring 1". Curve 2a (open
triangles), curve 2b (open triangles with a vertical bar), curve
2c (closed triangles) are the inductances $L_{1}(T)$, $L_{2}(T)$
and $L_{exp}(T)$ at $T=1.236$, and 1.243 K, respectively, for the
structure "ring 2". Curve 3a (open squares), curve 3b (open
squares with a vertical bar), curve 3c (closed squares) are the
inductances $L_{1}(T)$, $L_{2}(T)$ and $L_{exp}(T)$ at $T=1.243$,
1.253, and 1.271 K, respectively, for the structure "ring 3".}
\end{center}
\end{figure}

\begin{figure}
\begin{center}
\includegraphics[width=1\linewidth]{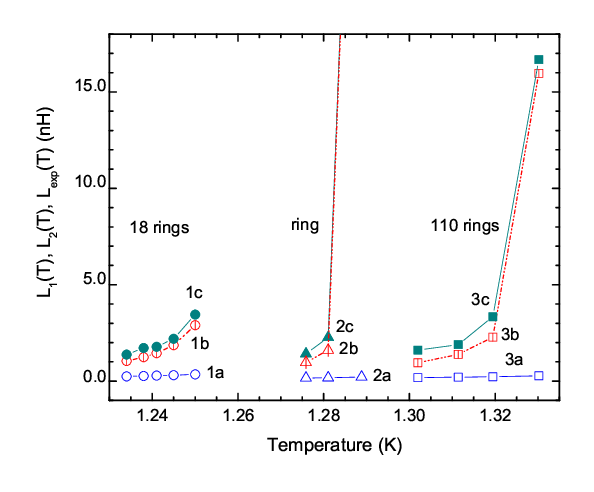}
\caption{\label{f9} (Color online) Curve 1a (open circles), curve
1b (open circles with a vertical bar), curve 1c (closed circles)
are theoretical inductances of wide, narrow semirings $L_{1}(T)$,
$L_{2}(T)$ and the experimental inductance of one ring
$L_{exp}(T)$ as a function of $T$ in the range 1.234-1.250 K,
respectively, for "18 rings". Curve 2a (open triangles), curve 2b
(open triangles with a vertical bar), curve 2c (closed triangles)
are the inductances $L_{1}(T)$, $L_{2}(T)$ and $L_{exp}(T)$,
respectively, for "ring" at $T=1.276$, 1.281, and 1.289 K. Data
2b, 2c at $T=1.289$ K are absent in the figure. These values are
given: $L_{2}(T=1.289)=46.6$ nH and $L_{exp}(T=1.289)=47.0$ nH.
Curve 3a (open squares), curve 3b (open squares with a vertical
bar), curve 3c (closed squares) are the inductances $L_{1}(T)$,
$L_{2}(T)$ and $L_{exp}(T)$ in the range 1.302-1.330 K,
respectively, for "110 rings".}
\end{center}
\end{figure}

In order to check the model of the experimental temperature
dependence of the phase shift $dl_{exp}(T)$ of the maxima of the
critical currents of different polarity with respect to the zero
flux in opposite directions along the flux axis and to clarify
adjustable parameters of the theoretical curves $dl(T)$ (Figs.
\ref{f6}, and \ref{f7}), we calculated the temperature dependences
of the inductances of wide and narrow semirings $L_{1}(T)$ and
$L_{2}(T)$, respectively, constituting the structures: "ring 1",
"ring 2", "ring 3" (Fig. \ref{f8}); "18 rings", "ring", "110
rings" (Fig. \ref{f9}). We compared the calculated total
inductance $L(T)=L_{1}(T)+L_{2}(T)$ with the experimental
inductance $L_{exp}(T)$ (Figs. \ref{f8}, and \ref{f9}) of an
individual ring from the structures under study.

In the limit $l(T)>>1$, the experimental inductance of one ring is
found from the expression \cite{clarke}
$L_{exp}(T)=\Phi_{0}/2I_{L}(T)$, where $I_{L}(T)$ is the maximum
value of the circulating current in the ring. The doubled value of
the circulating current is equal to the modulation depth of the
ring critical current
$2I_{L}(T)=dI_{c}(T)=I_{c}^{max}(T)-I_{c}^{min}(T)$, where
$I_{c}^{max}(T)$ and $I_{c}^{min}(T)$ are the maximum and minimum
values of the measured critical current at a certain temperature
in a small magnetic field.

The maximum and minimum values $I_{c}^{max}(T)$ and
$I_{c}^{min}(T)$ of the experimental critical current were taken
directly from the curves $I_{c+}(\Phi/\Phi_{0}+\phi_{l}(T))$ and
$I_{c-}(\Phi/\Phi_{0}-\phi_{l}(T))$ recorded in six
circularly-asymmetric aluminum structures at different
temperatures \cite{nikulov07, gurtvoi0910, gurtvoi0603005,
burlakov0609345}.

Inductances $L_{1}(T)$, $L_{2}(T)$ and $L_{exp}(T)$ are determined
at the same temperatures as  the temperature dependences of the
phase shifts $dl_{exp}(T)=2 \pi \phi_{l}(T)$ (Figs. \ref{f6}, and
\ref{f7}) of the maxima of the critical currents of different
polarity with respect to the zero flux in opposite directions
along the flux axis.

Figure \ref{f8} shows three sets of data, each consisting of two
theoretical relationships $L_{1}(T)$, $L_{2}(T)$ and one
experimental one $L_{exp}(T)$, for three structures: "ring 1",
"ring 2", "ring 3", measured in three temperature ranges. The
first set of data: curve 1a (open circles), curve 1b (open circles
with a vertical bar), curve 1c (closed circles) - presents the
inductances $L_{1}(T)$, $L_{2}(T)$ and $L_{exp}(T)$, respectively,
for "ring 1" at $T=1.211$, 1.216, 1.222, 1.227, 1.232, and 1.235
K.

Theoretical inductances of wide and narrow semirings for six
structures \cite{nikulov07, gurtvoi0910, gurtvoi0603005,
burlakov0609345} are given by the expressions
\begin{equation}
\label{e3}
L_{1}(T)=\mu_{0}\lambda(0)^{2}(1-T/T_{cf1})^{-1}l_{r}/(w_{w}d_{f}),
\end{equation}
\begin{equation}
\label{e4}
L_{2}(T)=\mu_{0}\lambda(0)^{2}(1-T/T_{cf2})^{-1}l_{r}/(w_{n}d_{f}),
\end{equation}
where it is taken into account that
$\lambda_{1}(T)=\lambda(0)(1-T/T_{cf1})^{-1/2}$,
$\lambda_{2}(T)=\lambda(0)(1-T/T_{cf2})^{-1/2}$.

The parameters of each structure such as: the length of the
semiring $l_{r}$, the width of the wide and narrow semirings
$w_{w}$, and $w_{n}$ are given earlier. The adjustable critical
temperatures $T_{cf1}$ and $T_{cf2}$ for the same structure are
the same as for the dependence $dl(T)$. The values of $\lambda(0)$
and the adjustable critical temperatures $T_{cf1}$ and $T_{cf2}$
are presented in the Table \ref{t1} for the structures: "ring 1",
"ring 2" and "ring 3". Since the thickness of the structures
\cite{nikulov07, gurtvoi0910, gurtvoi0603005, burlakov0609345} is
not indicated exactly, we used the thickness as an additional
adjustable parameter $d_{f}$.  Adjustable thickness values $d_{f}$
are shown in the Table \ref{t2} for each structure.

Curve 2a (open triangles), curve 2b (open triangles with a
vertical bar), curve 2c (closed triangles) present the dependences
$L_{1}(T)$, $L_{2}(T)$ and $L_{exp}(T)$, respectively, for the
structure "ring 2" at $T=1.236$, and 1.243 K.

Curve 3a (open squares), curve 3b (open squares with a vertical
bar), curve 3c (closed squares) present the functions $L_{1}(T)$,
$L_{2}(T)$ and $L_{exp}(T)$, respectively, for the structure "ring
3" at $T=1.243$, 1.253, and 1.271 K.

We found that the inductance of one ring, determined from the
experiment $L_{exp}(T)$ (curves 1c, 2c, and 3c), is close to the
calculated total inductance of the ring $L(T)=L_{1}(T)+L_{2}(T)$
for three structures: "ring 1", "ring 2", "ring 3" (Fig.
\ref{f8}). It can be seen that the main contribution to the total
inductance of the ring is made by the inductance of the narrow
semiring $L_{2}(T)$. The inductance $L_{1}(T)$ weakly depends on
$T$ and has an order of magnitude of 0.1 nH (curves 1a, 2a, and
3a). At the same time $L_{2}(T)$ strongly depends on $T$, since
$T$ is close to the critical temperature $T_{cf2}$ (curves 1b, 2b,
and 3b). The inductance $L_{2}(T)$ varies from 0.3 nH to 1.0 nH.

Figure \ref{f9} shows three sets of data, each of which consists
of two theoretical semiring inductances $L_{1}(T)$, $L_{2}(T)$ and
the experimental inductance of one ring $L_{exp}(T)$, for three
structures: "18 rings", "ring", "110 rings", measured in three
temperature intervals.

Curve 1a (open circles), curve 1b (open circles with a vertical
bar), curve 1c (closed circles) present the functions $L_{1}(T)$ ,
$L_{2}(T)$ and $L_{exp}(T)$, respectively, for the structure "18
rings" at $T=1.234$, 1.238, 1.241, 1.245, and 1.250 K. The
theoretical inductances of wide $L_{1}(T)$ and narrow $L_{2}(T)$
semirings for three structures are given by expressions similar to
expressions (\ref{e3}, \ref{e4}) used for three rings (Fig.
\ref{f8}).

The parameters of each structure are given earlier. The adjustable
critical temperatures $T_{cf1}$ and $T_{cf2}$ for the same
structure are the same as for $dl(T)$. The values of $\lambda(0)$
and the adjustable critical temperatures $T_{cf1}$ and $T_{cf2}$
are presented in the Table \ref{t1} for the structures: "18
rings", "ring", and "110 rings". Adjustable thickness values
$d_{f}$ are shown in the Table \ref{t2} for each structure. We
found that the total theoretical inductance
$L(T)=L_{1}(T)+L_{2}(T)$ differs from the experimental inductance
of one ring $L_{exp}(T)$ by ten percent, so that
$L(T)=0.9L_{exp}(T)$.

Curve 2a (open triangles), curve 2b (open triangles with a
vertical bar), curve 2c (closed triangles) present the dependences
$L_{1}(T)$, $L_{2}(T)$ and $L_{exp}(T)$, respectively, for the
structure "ring" at $T=1.276$, 1.281, and 1.289 K. Data 2b, 2c at
$T=1.289$ K are not shown in the figure. The values are given:
$L_{2}(T=1.289)=46.6$ nH and $L_{exp}(T=1.289)=47.0$ nH. We found
that $L(T)=0.75L_{exp}(T)$.

Curve 3a (open squares), curve 3b (open squares with a vertical
bar), curve 3c (closed squares) present the inductances
$L_{1}(T)$, $L_{2}(T)$ and $L_{exp}(T)$, respectively, for the
structure "110 rings" at $T=1.302$, 1.311, 1.320, and 1.330 K. We
found that $L(T)=(0.7-0.8)L_{exp}(T)$.

It can be seen that the main contribution to the total inductance
of the ring is made by the inductance of the narrow semiring
$L_{2}(T)$. The inductance $L_{1}(T)$ has an order of magnitude of
0.3 nH (curves 1a, 2a, and 3a). The inductance $L_{2}(T)$ can
reach 16 nH and 46.0 nH for curves 3b, 2b for the structures "110
rings" and "ring", respectively.

Thus, in order to confirm the model of the temperature-dependent
phase shifts of the maxima of the critical currents of different
polarity with respect to the zero flux in opposite directions
along the flux axis, we calculated the inductances $L_{1}(T)$ and
$L_{2}(T)$ of wide and narrow semirings and the experimental
dependence $L_{exp}(T)$ for one ring included in the structures
considered here. We found that the total theoretical inductance of
the wide and narrow semirings $L(T)=L_{1}(T)+L_{2}(T)$ is close to
the experimental inductance $L_{exp}(T)$ of one ring. In addition,
we improved the adjustable parameters: the critical temperatures
of the wide and narrow semirings $T_{cf1}$, and $T_{cf2}$ for the
theoretical curves $dl(T)$ (Figs. \ref{f6}, and \ref{f7}) and the
thickness of the structures $d_{f}$.

\section{Conclusion}

In this work, the oscillations of the rectified direct voltage
$V_{rec}(\Phi/\Phi_{0})$ are measured as a function of the
magnetic flux $\Phi/\Phi_{0}$ permeating a structure of three
circularly-asymmetric superconducting aluminum rings in series
biased with alternating current (without a dc component) at $T$
close to $T_{c}$. In an asymmetric structure, the rectified
voltage is due to the shifts $\phi_{l}(T)$ of the maxima of the
critical currents of different polarity
$I_{c+}(\Phi/\Phi_{0}+\phi_{l}(T))$ and
$I_{c-}(\Phi/\Phi_{0}-\phi_{l}(T))$ with respect to the zero flux
in opposite directions along the flux axis. Oscillations reach
maxima at $\Phi/\Phi_{0}=n+\phi_{l}(T)$ and minima at
$n-\phi_{l}(T)$, here $\phi_{l}(T)=0.16$ is the shift equal to the
shift of  the maxima of the critical currents. Oscillations take
zero values at $\Phi/\Phi_{0}=n$ and $n+1/2$.

In addition, ac voltage oscillograms $V_{ac}(\Phi/\Phi_{0},t)$ are
recorded at certain values of $\Phi/\Phi_{0}$ on two and three
rings in series biased with alternating cosine current (with a
zero dc component) with the frequency of 1.5333 kHz and an
amplitude close to the critical current $I_{c}(T,B=0)$.

Moreover, for the first time, we have sketched a model of the
temperature-dependent phase shift of the maxima of the critical
currents of different polarity with respect to the zero flux in
opposite directions along the normalized flux axis in a
circularly-asymmetric aluminum ring permeated with a magnetic
flux. It has been shown that this phase shift is equal to the
difference between the dimensionless kinetic inductances of the
wide and narrow semirings $dl(T)=l_{1}(T)-l_{2}(T)$ and arises
when the critical current densities are not the same in both
semirings having the same length and thickness. This is possible
only in a situation where the critical temperatures of the wide
and narrow semirings $T_{cw}$ and $T_{cn}$, respectively, differ.

We found that the temperature dependence $dl(T)$ is nonmonotonic.
At $T<T_{cn}$, both semirings are superconducting. The critical
current densities $j_{c1}(T)$ and $j_{c2}(T)$ in wide and narrow
semirings are the Ginzburg-Landau depairing current densities. In
this case, the phase shift
$dl(T<T_{cn})=l_{1}(T<T_{cn})-l_{2}(T<T_{cn})$ increases with
increasing temperature.

At $T_{cn}<T<T_{cw}$, the wide semiring is a superconducting,
while the narrow semiring is the Josephson S-s-S structure. The
critical currents of the wide and narrow semirings are determined
by the G-L depairing current and the Josephson critical current,
respectively. The phase shift
$dl(T_{cn}<T<T_{cw})=l_{1}(T<T_{cn})-l_{2}(T_{cn}<T<T_{cw})$
decreases with increasing temperature.

This model is used to theoretically describe the experimental
temperature dependence of the phase shift $dl_{exp}(T)=2 \pi
\phi_{l}(T)$.

In this model, the case of a large dimensionless inductance of one
ring $l(T)=l_{1}(T)+l_{2}(T)>>1$ is practically fulfilled at
experimental temperatures.

The experimental values of the shift $\phi_{l}(T)$ were taken
directly from the curves of magnetic-field-dependent critical
currents of different polarity $I_{c+}(\Phi/\Phi_{0}+\phi_{l}(T))$
and $I_{c-}(\Phi/\Phi_{0}-\phi_{l}(T))$, measured on six
asymmetric superconducting aluminum structures at different
temperatures \cite{karpii07, nikulov07, gurtvoi0910,
gurtvoi0603005, burlakov0609345}.

The adjustable critical temperatures of the wide and narrow
semirings $T_{cf1}$, and $T_{cf2}$ for each theoretical curve
$dl(T)$ and the adjustable thickness of each structure $d_{f}$
were refined after comparing the sum of the theoretical
inductances of the wide and narrow semirings
$L(T)=L_{1}(T)+L_{2}(T)$ and the experimental inductance
$L_{exp}(T)$ of an individual ring of the structure.

We have found a good agreement between the theoretical model and
measurements \cite{karpii07, nikulov07, gurtvoi0910,
gurtvoi0603005, burlakov0609345}. Note that we do not pretend to
provide a complete theoretical description of the
temperature-dependent phase shift, but only indicate the direction
in which to move.

In this model, we did not take into account the influence of the
magnetic field on the critical current density in wide and narrow
semirings, and, therefore, on the phase shift of the maxima of the
critical currents $dl(T)$, since the field values \cite{karpii07,
nikulov07} were much less than the maximum critical fields of
semirings.

Since the maximum critical fields of the semirings are large, the
amplitude from peak to peak oscillations of the rectified voltage
$V_{rec}(\Phi/\Phi_{0})$ on two rings in series (Fig. \ref{f2})
practically does not change in a certain interval of small fields.
Surprisingly, outside this range of fields, the amplitude of the
oscillations decreases sharply.

The model of the phase shift $dl(T)=2 \pi \phi_{l}(T)$ makes it
possible to explain the paradoxical fact that there is no shift
$\phi_{l}(T)$ of minima and maxima relative to integer and
half-integer values of the normalized flux on the dependence of
resistance oscillations $dR(\Phi/\Phi_{0})$ in a
circularly-asymmetric ring, while this shift is observed in the
dependences of critical currents with different polarity
$I_{c+}(\Phi/\Phi_{0}+\phi_{l}(T))$ and
$I_{c-}(\Phi/\Phi_{0}-\phi_{l}(T))$ for the same ring
\cite{karpii07}. We believe that this contradiction can be removed
due to the fact that $dR(\Phi/\Phi_{0})$ oscillations are usually
measured at a very small applied ac and a temperature
corresponding to the middle of the resistive N-S transition. With
these parameters, almost zero shift of the minima and maxima of
the resistance $dR(\Phi/\Phi_{0})$ is expected.

So, we have solved the long-standing difficult to explain
mysterious problem of the phase shift of the maxima in the
critical currents of different polarity with respect to the zero
flux in opposite directions along the normalized flux axis in a
circularly-asymmetric superconducting aluminum ring permeated with
a magnetic flux.

\section{Acknowledgments}
We are grateful to A. Firsov for making the structures. This work
was financially supported in the framework of STATE TASK No.
075-00355-21-00.



\end{document}